%% file: main.tex
\documentclass{article}
\usepackage[utf8]{inputenc}
\usepackage{authblk}
\usepackage{setspace}
\usepackage{graphicx}
\usepackage{float}
\usepackage{listings}
\usepackage{booktabs}
\usepackage{siunitx}
\usepackage{amsmath,amssymb}
\usepackage{fancyvrb}
\usepackage[super, comma, numbers, sort&compress]{natbib}
\makeatletter 
\renewcommand\@biblabel[1]{#1.} 
\makeatother

\usepackage{caption}
\DeclareCaptionFont{mysize}{\fontsize{8}{10}\selectfont}
\usepackage{color}
\usepackage[normalem]{ulem}
\definecolor{ultramarine}{RGB}{0,32,96}
\newcommand{\add}[1]{#1}
\newcommand{\delete}[1]{}

\newcommand{\comment}[2]{}

\usepackage{fancyvrb}
\usepackage{varioref,nameref,etoolbox}
\usepackage{hyperref}

\makeatletter 
\newcommand\mynobreakpar{\par\nobreak\@afterheading\vspace{0.1in}} 
\makeatother

\usepackage{framed}
\definecolor{shadecolor}{RGB}{248,248,248}

\definecolor{britishracinggreen}{rgb}{0.0, 0.26, 0.15}
\hypersetup{
	pdftitle={inferential Bland-Altman},
	pdfauthor={Siqueira \& Silveira},
	pdfsubject={Template},
	pdfkeywords={empty LaTeX},
	pdfpagemode=UseOutlines,
	bookmarksopenlevel=3,
    colorlinks=true,
    linkcolor=britishracinggreen,
    citecolor=britishracinggreen,
    urlcolor=britishracinggreen
}

\emergencystretch=\maxdimen
\title{Is Bland-Altman plot method useful without inference for accuracy, precision, and agreement?}

\date{\today}

\author[1]{Paulo Sergio Panse Silveira}
\author[2]{Joaquim Edson Vieira}
\author[3,*]{Jose Oliveira Siqueira}

\affil[1]{Department of Pathology, School of Medicine, University of Sao Paulo,  São Paulo, SP, Brazil - email: silveira@usp.br, ORCID: 0000-0003-4110-1038}
\affil[2]{Anesthesiology, Department of Surgery, School of Medicine, University of São Paulo, São Paulo, SP, Brazil - email: joaquimev@usp.br, ORCID: 0000-0002-6225-8985}
\affil[3]{Department of Pathology, School of Medicine, University of São Paulo, São Paulo, SP, Brazil - email: siqueira@usp.br, ORCID: 0000-0002-3357-8939}
\affil[*]{
    \raggedright
    {
    \textbf{Corresponding author}: \newline
    Av. Dr. Arnaldo, 455 \newline
    Cerqueira César, São Paulo - SP, 01246-903 \newline
    phone: +55 11 3061-7234
    email: siqueira@usp.br
    }
} 

\begin{document}

\maketitle

% comment this line to remove topic summary
% \tableofcontents

\include{Authors}

\include{Abstract}

\include{Introduction}

\include{Methods}
\include{Results}

\include{Discussion}

\bibliography{references} 
\bibliographystyle{plainnat}
% \bibliographystyle{dinat}
% \bibliographystyle{IEEEtran}
% \bibliographystyle{chicago}
% \bibliographystyle{vancouver}
% \bibliographystyle{unsrtnat}

% \include{Figures} % to leave all figures in the end

% % Appendix
% \pagenumbering{Roman}
% % restart figure numbering
% \renewcommand\thefigure{\arabic{figure}\tiny{s}}
% \setcounter{figure}{0}                   
% % restart equation numbering
% \renewcommand{\theequation}{\arabic{equation}\tiny{s}\normalsize}    
% \setcounter{equation}{0}  % reset counter
% % restart table numbering
% \renewcommand{\thetable}{\arabic{table}\tiny{s}\normalsize}    
% \setcounter{table}{0}  % reset counter
% \appendix
% \include{Appendix}

\end{document}

%% file: Authors.tex
\clearpage

\section*{Author’s full name, role, and affiliations}

\begin{itemize}
    \item Paulo Sergio Panse Silveira$^{1}$\newline
Conceptualization, Data Curation, Investigation, Methodology, Software, Validation, Visualization, Writing – Original Draft Preparation, Writing – Review \& Editing.\newline
ORCID: 0000-0003-4110-1038
    \item Joaquim Edson Vieira$^{2}$\newline
Conceptualization, Supervision, Validation, Writing – Review \& Editing.\newline
ORCID: 0000-0002-6225-8985
    \item José de Oliveira Siqueira$^{1,*}$\newline
Conceptualization, Data Curation, Formal Analysis, Methodology, Software, Validation, Writing – Review \& Editing.\newline
ORCID: 0000-0002-3357-8939
\end{itemize}

\subsection*{Affiliations}

\begin{enumerate}
    \item Medical Informatics, Department of Pathology, 
    \item Anesthesiology, Department of Surgery,
\end{enumerate}

Medical School at the University of São Paulo, Brazil.

%% file: Abstract.tex
\large{\textbf{Is the Bland-Altman plot method useful without
an inferential approach for accuracy, precision,
and agreement?}}
\normalsize

\section*{Abstract}

\textbf{Objective:} Bland and Altman plot method is a widely cited and applied graphical approach for assessing the equivalence of quantitative measurement techniques, usually aiming to replace a traditional technique with a new, less invasive, or less expensive one. Although easy to communicate, Bland and Altman plot is often misinterpreted by lacking suitable inferential statistical support. Usual alternatives, such as Pearson’s correlation or ordinal least-square linear regression, also fail to locate the weakness of each measurement technique. \textbf{Method:} Here, inferential statistics support for equivalence between measurement techniques is proposed in three nested tests based on structural regressions to assess the equivalence of structural means (accuracy), the equivalence of structural variances (precision), and concordance with the structural bisector line (agreement in measurements obtained from the same subject), by analytical methods and robust approach by bootstrapping. Graphical outputs are also implemented to follow Bland and Altman's principles for easy communication. \textbf{Results:} The performance of this method is shown and confronted with five data sets from previously published articles that applied Bland and Altman's method. One case demonstrated strict equivalence, three cases showed partial equivalence, and one showed poor equivalence. The developed R package containing open codes and data are available with installation instructions for free distribution at Harvard Dataverse at https://doi.org/10.7910/DVN/AGJPZH. \textbf{Conclusion:} It is possible to test whether two techniques may have full equivalence, preserving graphical communication according to Bland and Altman's principles, but adding robust and suitable inferential statistics. Decomposing the equivalence in accuracy, precision, and agreement helps the location of the source of the problem in order to fix a new technique.

\section*{Descriptors}

Measurement Techniques, Equivalence Testing, Confidence Intervals, Statistical Inference, Statistical Graphics, Statistical Analysis, Applied Statistics, Software Packages, Regression Analysis

%% file: Introduction.tex
\section{Introduction}\label{sec:introduction}

Bland and Altman's~\cite{Bland1986} paper introduced a graphical approach to compare two measurement techniques using peak flow meters, which has become well-known and widely used in various medical fields. This method has been applied to compare modern peak flow meters~\cite{Pesola2010}, DNA sequencing methods~\cite{Misyura2018}, athletes' performance~\cite{Atkinson1998}, blood pressure measurements~\cite{Shimada2015}, muscle tone quantification~\cite{Lo2017}, and validation of self-reported height and weight~\cite{Aasvee2015}. It has been referenced in over 35,000 scientific publications.

In a nutshell, Bland-Altman plots assess the 95\% limit of agreement (LoA) given by a band from the mean difference~$\pm 1.96$~standard deviation of two techniques' measurements. If the range between the lower and upper LoA is clinically unimportant, the techniques are assumed equivalent~\cite{Jones2011,Taffe2020,Parker2020}. More recently, confidence intervals were added at the upper and the lower LoA~\cite{Creasy1956,Zou2013,Carkeet2015,Taffe2020b,Christensen2020} to provide some range for tolerance. However, this tolerance only provides a statistical test for the band limits, not an additional decision for technique equivalence. Bland-Altman plot method is, therefore, subjective~\cite{Watson2010}. The clinical importance is attributable by the researcher as a threshold and it is a situation equivalent to acceptance of a null hypothesis by visual inspection of the graph without any measurement of the amount of equivalence and inferential statistical support.

Due to a lack of statistical support, the equivalence approach led to misunderstandings and anecdotal interpretation of data, sometimes contrary to the original author's recommendation. It is often misinterpreted as ``two exams are equivalent when the majority of data are within the band limits,"~\cite{Watson2010,Giavarina2015} which is always true, ranging from 75\% to 100\% independently of the data distribution according to Chebychev's inequality theorem~\cite{Frost,Savage1961}, or as ``the points inside the band must be uniformly distributed", which was never stated by the original authors. Bland-Altman plot method is insufficient as it only provides a visual decision.

\add{Although widely used, the Bland-Altman plot method lacks a clear null hypothesis on method equivalence and, consequently, cannot have statistical decision-making and relies on subjective judgment through visual inspection~\cite{Watson2010}. The available packages in the R language are not sufficiently clear or do not provide a comprehensive solution to determine when two measurement techniques can be considered equivalent.} 

The present work applies a three-step statistical decision allowing the researcher to determine if there are enough elements to reject the equivalence of the two techniques. The solution applies three nested tests with $p$ values and robust statistical decisions by bootstrapping. This method was implemented in a freely-distributable R~package and the whole analysis, including statistics and graphical outputs, requires a single command line to be executed by the researchers.

%% file: Methods.tex
\section{Methods}\label{sec:methods}

This investigation proposes the addition of statistical criteria to Bland and Altman's plot method~\cite{Bland1986}. Since it is a purely theoretical approach, it was not submitted to any ethics committee. 

The R package containing open codes and sample data is available with installation instructions on Harvard Dataverse for free distribution~\cite{eirasBA}.

\subsection*{Rationale}

Three steps to claim strict equivalence between measurement techniques are proposed, respectively checking (1)~equivalence of structural means (equality of accuracy), (2)~structural variances (equality of precision), and (3)~agreement with the structural bisector line (equal measurements obtained from the same subject). Full equivalence may be assumed when there is non-rejection of equivalence in all three tests. The significance level of 5\% was adopted in this text. 

At first, the statistical approach may seem somewhat convoluted because researchers have only observed data, while decisions depend on structural, non-observable values. The obscure term `structural' in this context refers to true values, estimated from a statistical approach necessary to purge observed measures from measurement errors~\cite{Isaac1970,Thoresen2007}. 

Regressions applied to all three tests are not crude regressions, but statistical artifices that connect structural values with functional procedures. The three tests involve statistical regressions that connect structural values with functional procedures, providing conclusions on accuracy, precision, and agreement. This approach combines scattered statistical theoretical results from 1879 to 2015~\cite{Creasy1956,Hedberg2015,Shukla1973,Glaister2001,Oldham1962,Linnet1998,Kummell1879,Albert1992}. \add{These tests are conceptually nested and propose inference based on solid mathematical foundation. The final test, which is also the most important one, assesses agreement with the bisector, demonstrating the reliability of the values obtained from two measurement techniques applied to each individual. This test depends on Deming regression~\cite{Creasy1956,Watson2010}, whose basic theorem was developed over a century ago~\cite{Kummell1879}. However, it would not make sense to verify such agreement if the two methods did not measure with equal precision, which is test 2 based on the theorem demonstrated by Shukla in 1973~\cite{Shukla1973}, and without the same accuracy, introducing a bias, which is test 1 based on Hedberg and Ayers in 2015~\cite{Hedberg2015}.}

Bootstrapping~\cite{Efron2007} is also used to compute confidence intervals in addition to analytical tests. It is shown in graphics to support the researcher's interpretation and to make it easier to communicate results. In our application, bootstrapping is represented by shadowed areas containing 95\% of all resampled regressions, which is assumed as the area containing the true populational regression.

The main concepts, ballasting the connection between structural null hypotheses and their functional correspondences, are outlined in the following. 

\subsection*{Observed and true variable values}

Measurements provided from a reference technique $A$ and candidate under assessment technique $B$ (each technique was applied once to each subject), according to the physics error theory, give: \mynobreakpar

\begin{eqnarray}
\label{eq:errors}
\begin{aligned}[b]
B: y = Y + \delta\\
A: x = X + \epsilon
\end{aligned}
\end{eqnarray}

where

$y$ and $x$ … are independent pairs of observed measurements,

$Y$ and $X$ … are the true correspondent measurements,

$\delta$ and $\epsilon$ … are independent measurement errors with null average.

These error terms appear because all measurement techniques have a certain degree of imprecision. Assuming that $Y$ and $\delta$, and $X$ and $\epsilon$ are also statistically independent and that these errors have no preferential direction (null averages, $E[\delta]=0$ and $E[\epsilon]=0$), the mean of all observed values is equal to the mean of true values ($\bar{y}=\bar{Y}$ and $\bar{x}=\bar{X}$), demonstrated by the respective expected ($E$) values:

\begin{eqnarray}
\label{eq:errorsdemo}
\begin{aligned}[b]
E[y] = {E[Y+\delta] = E[Y] + E[\delta]} = E[Y]\\
E[x] = {E[X+\epsilon] = E[X] + E[\epsilon]} = E[X]
\end{aligned}
\end{eqnarray}

Consequently, the observed mean difference between techniques is also equal to the structural bias ($\bar{y}-\bar{x} = \bar{Y}-\bar{X}$). These equalities allow the correspondence between functional computation and structural hypotheses, reducing all three nested tests to two ordinary least square linear regressions and one Deming regression. The relationship between structural and functional tests follows.

\subsection*{Test 1: Accuracy}

Hedberg and Ayers applied a covariate with measurement error in the analysis of covariance (ANCOVA) in order to test mean structural equality for these repeated measure designs~\cite{Hedberg2015}. This simple linear regression applies the differences between measurements obtained from the same subjects, $y_i-x_i$, and the centered value of the reference measurement, $x_i-\bar{x}$. 

The null hypotheses:
\begin{eqnarray}
\begin{aligned}[b]
\text{structural}~ H_{0,1}: & ~E[X]=E[Y]\\
\text{functional}~ H_{0,1}: & \alpha=0\\
\text{computing the regression}~ & ~y_i - x_i = \alpha + \beta (x_i-\bar{x}) + \nu_i
\end{aligned}
\end{eqnarray}
in which $\nu_i$ is the error term.

By centering values on the $x$ axis, by subtracting $\bar{x}$ from each original value $x_i$, the intercept, $\alpha$, of a regression line becomes more meaningful as it corresponds to the mean of $y-x$, while the slope, $\beta$ is not affected. By this artifice, this allows for the assessment of the equivalence of means of measurements obtained from different techniques, with the intercept representing the mean difference. Analytically, the null hypothesis of no mean difference is not rejected when zero is in the 95\% confidence interval of the intercept.

Graphically, the regression intercept is the mean of $y-x$ and located where the line crosses the $y$ axis. The null hypothesis is $(0,0)$, meaning no difference between techniques. If bootstrapping shows $(0,0)$ outside the 95\% confidence interval, the null hypothesis is rejected.

\subsection*{Test 2: Precision}

Verification of equal variability of measurement errors obtained from two techniques is based on Shukla~\cite{Shukla1973} and also independently adopted by Oldham~\cite{Oldham1962} without widespread application. The null hypotheses are:
\begin{eqnarray}
\begin{aligned}[b]
\text{structural}~ H_{0,2}: & ~\lambda=\frac{V[\delta]}{V[\epsilon]}=1\\
\text{functional}~ H_{0,2}: & ~\rho(x-y,x+y)=0\\
\text{computing the regression}~ & ~y_i - x_i = \alpha + \beta (x_i+y_i) + \theta_i
\end{aligned}
\end{eqnarray}
in which $\theta_i$ is the error term.

The structural null hypothesis computes lambda as the ratio between the variance of measurement errors; if the variability of errors is similar ($\lambda=1$) then the precisions of both techniques are similar. 

It was demonstrated that a regression of $y-x$ against $x+y$ can detect unequal precisions, as the slope of the regression will not be null when the true value of ${\lambda \ne 1}$~\cite{Shukla1973,Shoukri2010}. Analytically, the null hypothesis of equal precisions is rejected if a horizontal line cannot be fitted in the 95\% confidence band defined by the functional regression. Graphically, rejection of this null hypothesis corresponds to the impossibility to fit a horizontal line in the 95\% confidence band defined by the functional regression. Note that when each technique is applied to each subject more than once, it requires correction for computing $\lambda$, which was implemented according to the NCSS Manual~\cite{NCSS}. Also, the axes proposed by Shukla are the same ones used by Bland and Altman's original concept~\cite{Bland1986}, showing that the original method only compares the precision between measurement errors and is not a full equivalence test.

\subsection*{Test 3: Bisector line agreement}

This test applies Deming regression to verify if two measurement techniques measure the same values in the same subjects~\cite{Creasy1956,Glaister2001,Linnet1998,Kummell1879,Albert1992}. While ordinary least square regression treats independent variable $x$ free of measurement error, Deming regression reasonably takes into account measurement error in both measurement techniques. Linnet~\cite{Linnet1998} studied several regression methods, showing that the Deming regression method is robust and performs better than ordinary least square regression.

When true values measured by two techniques coincide, ordered pairs of these measures follow the true bisector line. Therefore, the null hypotheses are:
\begin{eqnarray}
\begin{aligned}[b]
\text{structural}~ H_{0,3}: & ~Y=X ~or~\\
                            & ~Y-X=\alpha+(\beta-1)X, ~where~ \alpha=0 ~and~ \beta=1\\
\text{functional}~ H_{0,3}: & ~E[y]=x\\
\text{computing the regression}~ & ~y_i - x_i = \alpha + (\beta-1)x_i + \delta_i - \beta\epsilon_i
\end{aligned}
\end{eqnarray}
in which $\delta_i$ and $\epsilon_i$ are error terms.

Deming regression is a method used to compare two measurement techniques, taking into account the errors in both techniques. Contrary to the regular ordinary least square regression statistical testing $\beta=0$, Deming regression verifies if the slope of the regression line is equal to 1 ($\beta=1$), which represents the bisector line agreement. In addition, $\beta$ simultaneously appears as the regression slope multiplying $x$ and as part of the regression overall error term ($\delta - \beta \epsilon$); transitively, it implies that $x$ becomes correlated with the combined error, preventing the computation of an ordinary least square regression~\cite{Isaac1970,Thoresen2007,Antonakis2010,McCartin2003,Roberts2012}. 

Deming regression also depends on $\lambda$, which was estimated in the previous step, to compute the true values $X$ and $Y$ before the computation of the regression estimates. When the value of lambda is not assumed to be 1, it affects the band width. Analytically, the null hypothesis is rejected if ${\alpha \ne 0}$ and ${\beta \ne 1}$. Since these two parameters are jointly estimated, Bonferroni correction is applied to control the probability of type I error to preserve test power (effective significance level is 2.5\%). Graphically, two alternative statistical approaches were implemented for bisector line agreement: the assessment of the 95\% prediction ellipse by bootstrapping and the assessment 95\% confidence band of regression by bootstrapping. Respectively, the null hypothesis is rejected if (${\beta,\alpha}$) is not inside the ellipse or the bisector line cannot be accommodate inside the band. These methods jointly test intercept and slope, and provide stronger statistical power than the independent appraisal of intercept and slope.

\subsection*{Translations}

The three tests are conducted using both analytical (based on $p$ value) and graphical (based on bootstrapping) approaches. In some cases, the analytical approach indicates no rejection of the null hypothesis while the graphical approach shows lines outside the confidence bands for precision and bisector line agreement tests due to differences in accuracy. This discordance can be attributed to bias in a particular technique, as seen in the example of Figure~\ref{fig:case2a}~(left panel). Therefore, a combination of analytical and graphical approaches is necessary for better interpretation of precision and agreement~\cite{Anscombe1973}, even in the presence of biased means.

The bias in accuracy can be corrected by translating the lines by the amount of bias computed. This correction enables the analytical approach to align with the graphical approach by positioning the lines inside the confidence band obtained by bootstrapping. Specifically, for a precision test, the null hypothesis is not rejected when a horizontal line shifted by the bias can be located inside the 95\% confidence band (as shown in the example of Figure~\ref{fig:case2a}, central panel). Similarly, in the bisector agreement test, non-rejection of the null hypothesis occurs when lines parallel to the bisector line translated by the bias range can be located inside the 95\% bootstrapping confidence regression band (as shown in the example of Figure~\ref{fig:case2a}, right panel).

%% file: Results.tex
\section{Results}\label{sec:results}

We revisited five data sets: from the original Bland and Altman data~\cite{Bland1986} (case 1), three other from Bland and Altman~\cite{Bland1999} (case 2), and one from data provided by Videira and Vieira~\cite{Videira2011} (case 3).

\subsection*{Case 1}

\begin{figure}[h!]
\begin{center}
\includegraphics[width=4.7in]{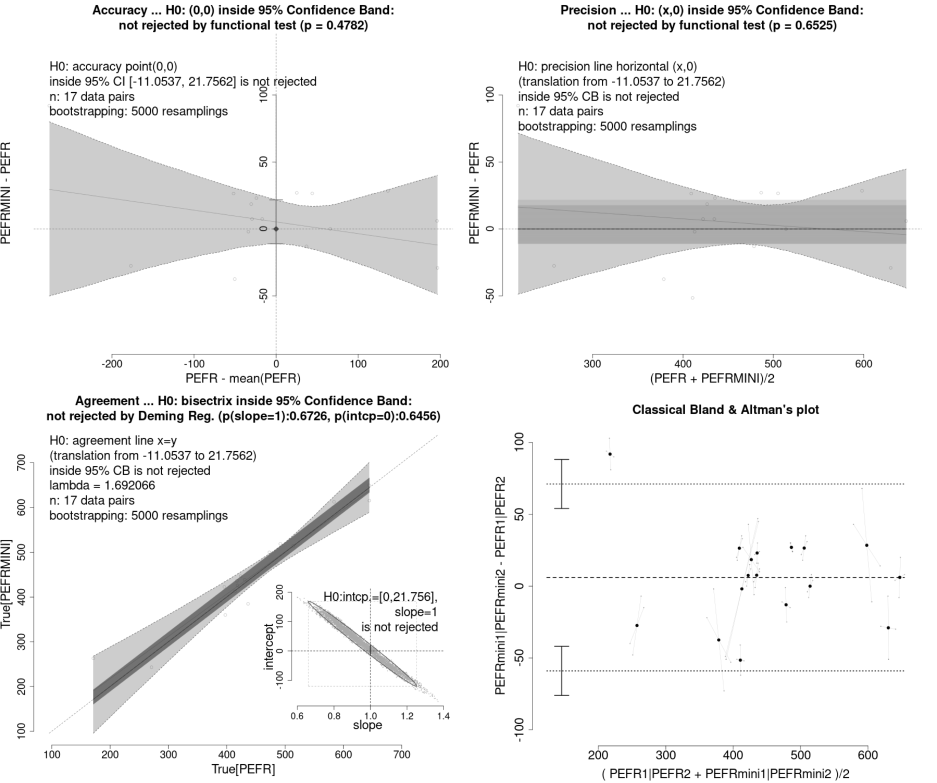}
\end{center}
\caption{Graphical representation from accuracy, precision, and bisector concordance tests showing that peak flow measurements from Wright and Mini PEFR are strictly equivalent. See text, case 1. A traditional Bland-Altman plot is depicted for comparison with the precision test.}\label{fig:case1}
\end{figure}

Bland and Altman proposed a graphical plot method to assess the equivalence of two peak expiratory flow rate (PEFR) measurement techniques, the Wright Peak Flow and Mini Wright Peak Flow meters. The study involved 17 subjects and both instruments were considered strictly equivalent. Figure~\ref{fig:case1} displays several statistical tests, including accuracy, precision, and bisector concordance, with $p$-values of 0.4782, 0.6525, 0.6726, and 0.6456, respectively. The structural regression bands were obtained by bootstrapping. The results show that the null hypothesis is inside the 95\% confidence interval for accuracy, within the 95\% confidence band defined by the structural regression for precision, and accommodated into the 95\% confidence band defined by Deming regression for bisector concordance ($\lambda=1.692$). Additionally, the right panel shows the 95\% prediction ellipse, an alternative way to jointly test slope and intercept. A traditional Bland-Altman plot was also included for comparison; note that the axes used are the same as those used in the agreement test.

\subsection*{Case 2}

Bland \& Altman provided other three application examples of their graphical method~\cite{Bland1999}. 

\begin{figure}[h!]
\begin{center}
\includegraphics[width=4.7in]{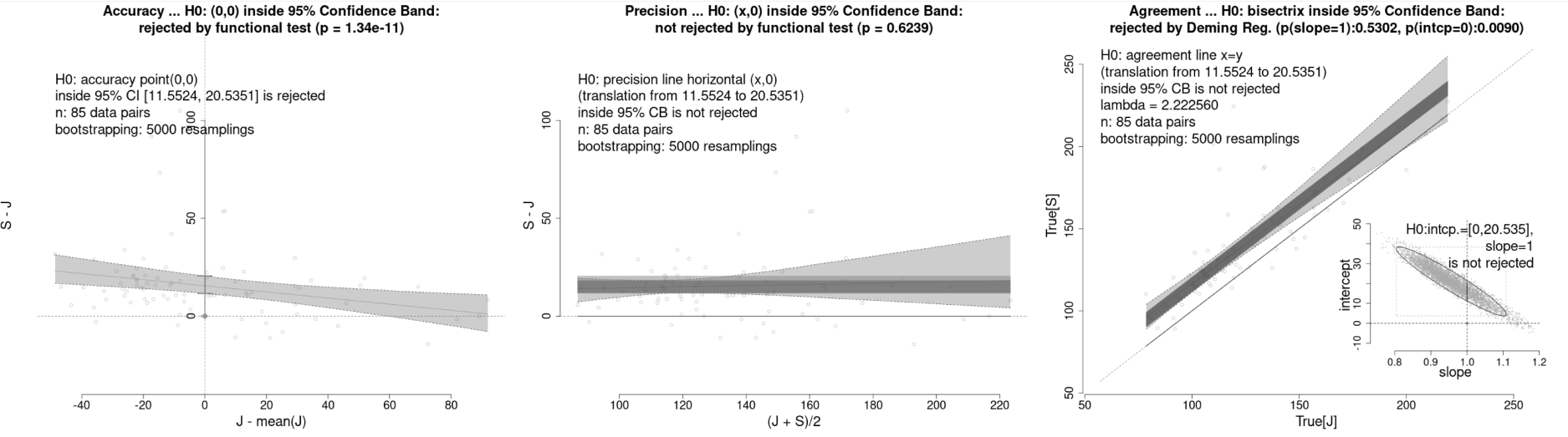}
\end{center}
\caption{Comparison of systolic blood pressure measured by a human observer J and an automatic machine S showing a structural bias (overestimation by S) at accuracy test, and concordance at the precision test and bisector test.  See text, case 2(a).}\label{fig:case2a}
\end{figure}

(\textbf{a}) In a comparison between systolic blood pressure measurements taken by an observer and an automatic machine, a systematic bias towards the machine was detected ($n=85$). While the authors concluded that equivalence could not be assumed due to a large interval range, our analysis showed that the observer and machine may be interchangeable after discounting the bias. The structural bias is represented by the 95\% confidence interval above the diamond, but precision and bisector line agreement tests were passed. The intercept is inside the 95\% prediction ellipse, and the non-null intercept cannot be corrected by traditional analytical approaches. (Figure~\ref{fig:case2a})

\begin{figure}[h!]
\begin{center}
\includegraphics[width=4.7in]{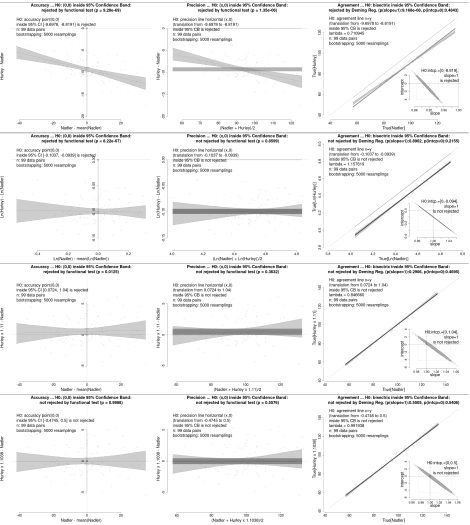}
\end{center}
\caption{Comparison of the percentage of plasma volume in blood provided by two different equations (Nadler and Hurley methods) using raw data (upper panels), logarithm transformation (second row), Hurley measurements x 1.11 (third row), and Hurley measurements x 1.1038. See text, case 2(b).}\label{fig:case2b}
\end{figure}

(\textbf{b}) The second example compares two techniques, Nadler and Hurley, for estimating the percentage of plasma volume in blood ($n=99$). The original authors found increasing bias towards Nadler's technique with greater average values. To verify equivalence, two strategies were proposed: logarithm transformation and scaling Hurley multiplied by 1.11. Figure~\ref{fig:case2b} shows our approach, which confirms no equivalence between methods in any of the three tests~(Figure~\ref{fig:case2b}, upper row). Logarithm transformation does not solve structural bias, but leads to equivalences in precision and agreement line~(Figure~\ref{fig:case2b}, second row). The multiplication of Hurley values by 1.11 is a more successful strategy, with marginal failure for accuracy~(Figure~\ref{fig:case2b}, third row). Using our approach, we found strict equivalence by multiplying Hurley values by approximately 1.1038, with improved precision and agreement line tests~(Figure~\ref{fig:case2b}, lower row).

\begin{figure}[h!]
\begin{center}
\includegraphics[width=4.7in]{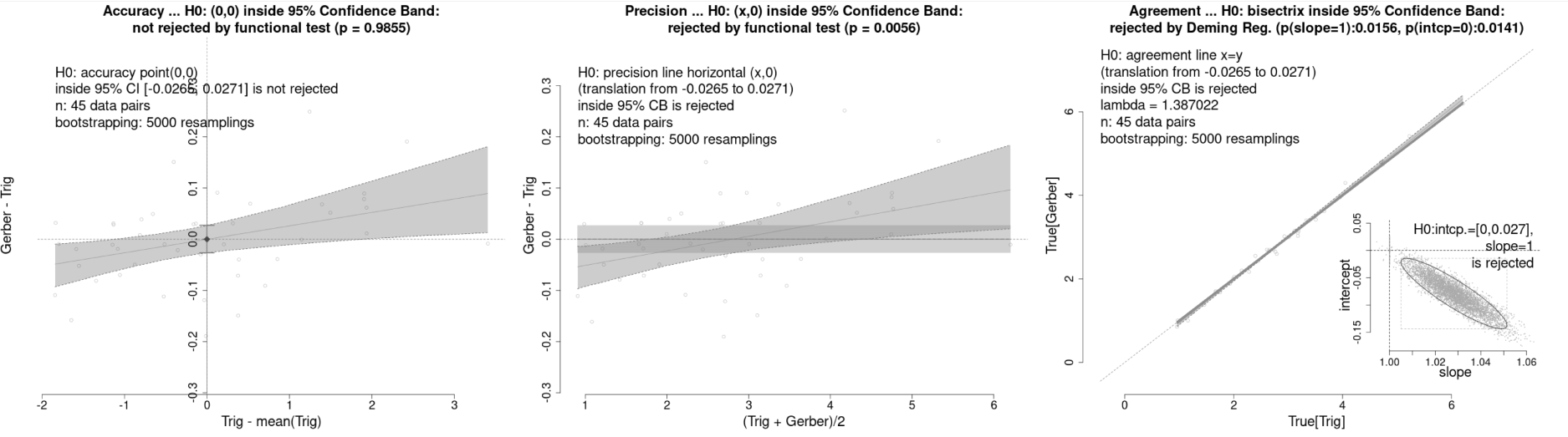}
\end{center}
\caption{Comparison of the content of fat in human milk measured by glycerol released by enzymic hydrolysis of triglycerides (Trig) and by the standard Gerber method. These two methods measured equal average (left panel), but are not strictly equivalents in precision (central panel) or by bisector concordance (right panel). See text, case 2(c).}\label{fig:case2c}
\end{figure}

(\textbf{c}) Bland and Altman compared fat content in human milk ($n=45$) using enzymic hydrolysis of triglycerides and by the standard Gerber technique. They found that one technique overestimated for smaller and underestimated for greater values, requiring adjustment of their traditional lines into a slanting band formed by two straight lines in order to accommodate these differences. Our proposal (Figure~\ref{fig:case2c}) naturally produced a slanted band, making the adjustment unnecessary. Our results contradict the authors' conclusion that the two techniques are equivalent in precision and agreement.

\subsection*{Case 3}

\begin{figure}[h!]
\begin{center}
\includegraphics[width=4.7in]{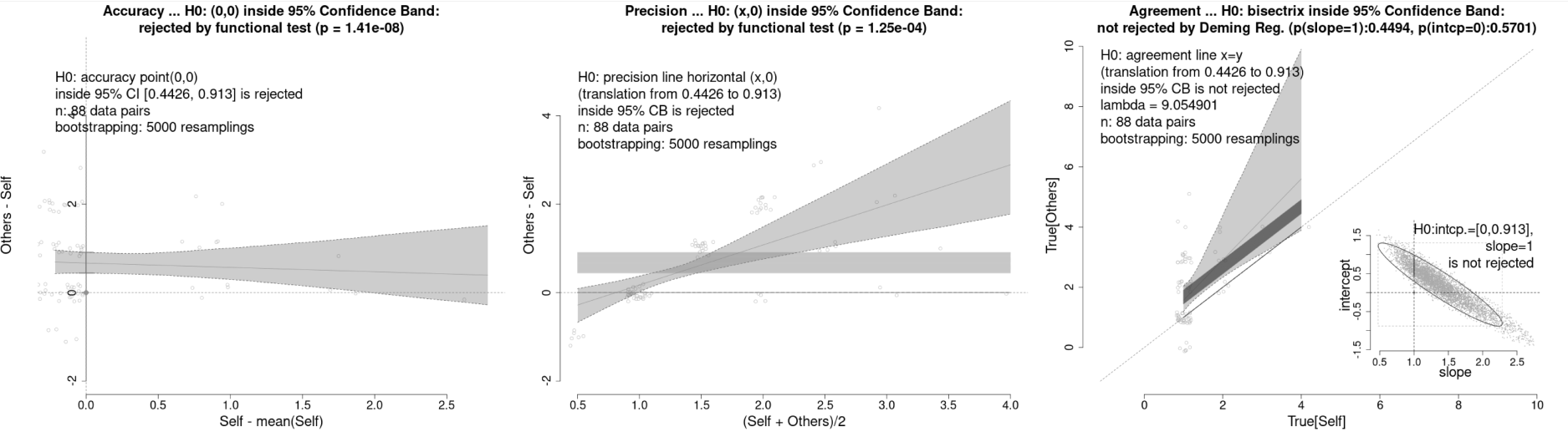}
\end{center}
\caption{Graphical representation of anesthesiologists' self-perception and peers' perception about their skills in deciding on the use of neuromuscular blocking drugs showing no equivalence in accuracy and precision. See text, case 3.}\label{fig:case3}
\end{figure}

Videira and Vieira~\cite{Videira2011} compared anesthesiologists' self-perception and their peers' perceptions of skills in deciding on the use of neuromuscular blocking drugs ($n=88$) using questionnaires. They found that self-perception and peer perception did not match, with subjects overestimating their abilities compared to their colleagues. Our approach (Figure~\ref{fig:case3}) supports this bias as the "above-average effect" (tendency to consider oneself better qualified) and also shows that the two perceptions are not equivalent. 

%% file: Discussion.tex
\section{Discussion}\label{sec:discussion}

Bland and Altman's analysis emphasizes clinical significance and their plots largely ignore statistical inference, relying on visual inspection to draw what is considered by Watson and Petrie as subjective conclusions~\cite{Watson2010}. Our contribution adds an objective statistical inference and locates causes of non-equivalence by taking apart accuracy, precision, and bisector agreement, but we followed the original Bland and Altman idea by preserving graphical outputs that make the communication easier~\cite{Parker2020}.

Altman and Bland argued that ``the use of correlation is misleading" and insufficient for comparing clinical measurements~\cite{Altman1983}, and also emphasized that ``comparability of techniques of measurement is an estimation problem: statistical significance is irrelevant"~\cite{Altman1986}. We respectfully disagree from the latter statement for it is necessary to compare related measurement techniques with proper confidence intervals. In fact, we look for statistical treatment comparing any two related measurement techniques and a proper method to compute confidence intervals instead of non-informative Chebychev's intervals with or without LoA additional flexibilization or adaptations to create slanted limits of agreement that these authors erroneously proposed~\cite{Bland2003}. 

In this study, we analyzed five published data sets using the Bland and Altman plot method. Our proposed three-step tests added statistical support and can locate the source of non-equivalence between techniques. For example, we found that in the case of peak flow expirometers~\cite{Bland1986}, there was strict agreement in accuracy, precision, and agreement line. The other three data sets~\cite{Bland1999} are examples of solvable equivalence between methods, but it was shown that our three-step tests provide solutions more easily. Finally, for the data set in Videira et al. (2011)~\cite{Videira2011}, our approach showed the conceptual importance of nested tests, as correcting the bias alone for the mean difference would result in a meaningless decision from the bisector line (third test) due to discrepancies in precision (second test). Without taking into account the nesting nature of our approach, one could have accepted equivalence despite the differences in precision.

\add{There are other packages that address the Bland-Altman plot method in R: blandr, MethComp, MethodCompare, and mcr. The blandr~\cite{blandr} package provides various ways to display the traditional plot, including limits of agreement (LoA) confidence intervals. Similar to the original method, the decision is solely based on visual inspection, and there is no Deming regression, which we consider the most fundamental part for assessing complete equivalence between two measurement methods. Interestingly, the example in the function blandr.method.comparison at the blandr package itself states that ``Paired T-tests evaluate for significant differences between the means of two sets of data. It does not test agreement, as the results of a T-test can be hidden by the distribution of differences," ``Correlation coefficients only tell us the linear relationship between 2 variables and nothing about agreement," and ``Linear regression models are conceptually similar to correlation coefficients, and again tell us nothing about agreement." However, despite these correct statements about the limitations of these statistics, all three insufficient statistics are still computed in the example.}

\add{MethodCompare~\cite{MethodCompare} is a package with a small set of functions that aims to compare bias and precision. The author of this package, Patrick Taffé, was cited in this text~\cite{Taffe2020,Taffe2020b}, and his aspiration is focused on improving the confidence intervals of the limits of agreement. Respectfully, in our opinion, works that focus on LoAs~\cite{Creasy1956,Zou2013,Carkeet2015,Taffe2020b,Christensen2020} do not address the fundamental issue of equivalence between methods. They are improvements on a secondary aspect of the Bland-Altman plot method that already faces the problem of dealing with an uninformative interval, as discussed below.}

\add{MethComp~\cite{MethComp} implements several maneuvers accumulated in the literature in an attempt to improve the Bland-Altman plot method. This package uses Passing and Bablok regression (PBreg), a non-parametric regression that suffers from the same problem as OLS regression by not taking into account the measurement error of one of the variables and, therefore, is also not the appropriate solution. This package has several functions with a large number of parameters. It is evident that the author invested a lot of time and care in their package, but there are no obvious tests related to verifying the bias or precision of the methods, which are necessary assumptions for Deming regression. Although this package includes Deming regression, it does not display it with confidence bands and focuses on comparing it with OLS regression, which (as the original authors of the Bland-Altman plot method themselves state~\cite{Altman1983}) should not be chosen. Among its parameters, it requests the value of lambda, with a default of 1, but does not provide resources to estimate it. 
Additionally, the estimation of the intercept and slope is not done jointly, which may incorrectly lead to the non-rejection of equivalence. This issue can be illustrated by the second row of Figure~\ref{fig:case2b}. You will notice a dotted, small rectangle drawn around the elliptical region of the agreement test. If the slope and intercept are not considered jointly, any point within this rectangle (i.e., any slope within the left and right limits, and any intercept within the bottom and top limits of this rectangle) would result in a statistical decision of non-rejection of the null hypothesis. However, the correct decision should only be made based on non-rejection when the point (1,0) is within the elliptical region. In the given example, non-rejection was only observed due to translation. However, it is important to note that without considering translation, MethComp would not be able to detect that the Deming regression line was not coincident with the bisector.}

\add{We have identified the package mcr~\cite{mcr} as the closest package to ours. However, it also has limitations in its approach. The package includes a simple Bland-Altman plot without Limits of Agreement (LoA) using the function plotDifference. The mcr package also includes Deming regression with the function mcreg. However, similar to MethComp package, it suffers from the issue of using a default lambda value of 1 without providing guidance on its estimation. The approach to bias in the mcr package is somewhat incomplete and misguided. It uses the observed bias ($y-x$) as the dependent variable and the observed values of the reference measurement as the independent variable, without centering it by its mean value ($x-\bar{x}$). Then, as shown in the examples of the function plotBias, it applies several variations of Deming regression, which should be performed with latent, true values, instead of observed values, with the aggravation of lacking a clear statistical test for decision-making. The documentation of this function and its regression variants is obscure. Apparently, the author is not aware of the theoretical foundation of Hedberg and Ayers~\cite{Hedberg2015}, which could have been used to develop a statistical test for accuracy. Furthermore, there are no functions that verify the equivalence of precisions, which is more important and more difficult to address in practice than the bias between a surrogate method and a reference method.}

\add{To our knowledge, this is the first time that a single procedure brings together and applies the results of Hedberg and Ayers~\cite{Hedberg2015}, Shukla~\cite{Shimada2015}, Shoukri~\cite{Shoukri2010}, and Linnet~\cite{Linnet1998}, providing a theoretical basis for the statistics related to accuracy, precision, and Deming regression, respectively. Additionally, analytical methods, bootstrapping, and easily interpretable graphical outputs have been implemented. Above all, although each function can be used independently (examples are detailed in the package documentation), we have created a coordinating function that allows researchers, with just an Excel or similar file containing their data in a data frame, to use a single command to generate a complete report in plain text, HTML, or PDF format. For example, the elements in Figure~\ref{fig:case1} were extracted from the report generated with the following command:}

\begin{scriptsize}
\begin{Verbatim}[frame=single]
out <- eirasBA::all.structural.tests(eirasBA::PEFR,
                              reference.cols=c(1:2),
                              newmethod.cols=c(3:4),
                              alpha=0.05, out.format="html")
\end{Verbatim}
\end{scriptsize}

\add{In addition, eirasBA provides treatment for repeated measures, which is not found in the other mentioned packages. It is common for researchers to take multiple measurements using the same technique when attempting to compare a new method they intend to use as a replacement for an established one. Depending on whether unique or repeated measures are provided, eirasBA calculates the value of lambda and automatically uses it in subsequent tests. This feature enhances the package's ability to handle repeated measurements effectively.}

\add{One of the most significant criticisms of both the traditional Bland-Altman plot method and the discussed packages is the reliance on visual inspection for decision-making. In this regard, eirasBA brings innovation by automating the recognition of lines or points within the regions of bands or ellipses, providing decision indicators to the users. This eliminates the subjective aspect of visual interpretation and enhances the objectivity of the decision-making process.}

\add{Another innovative concept of line translations was introduced, allowing the assessment of precision and bisector agreement even in the presence of unequal means between two measurement techniques. Biased techniques that provide equal precision and agreement may still be useful with a simple calibration or correction. Reversely, if a surrogate technique is unbiased but less precise, it could be eligible as a screening step; however, if this imprecision imposes risks to patients, then the technique must be reviewed. In essence, the decomposition of accuracy, precision, and agreement with bisector line analysis can guide researchers in determining where to focus their efforts to improve a new technique when full equivalence is not achieved.}

\add{A noteworthy observation is that the axes used in the precision test are the same as those used in the original Bland-Altman plot method. Contrary to the belief of many users, the original method, even under optimal conditions, reflects a comparison of precision rather than equivalence between two measurement techniques. However, even this comparison is not possible because the original Bland-Altman bands (i.e., the so-called limits of agreement, LoA) do not represent a confidence interval and merely correspond to the limits of a Chebyshev interval~\cite{Frost, Savage1961}. Chebyshev's inequality theorem provides information about the percentage of data that is guaranteed to fall within a given interval, regardless of the probability distribution. For instance, in a normal distribution, approximately 95\% of the data falls within plus or minus two standard deviations around the mean, while a minimum of 75\% is guaranteed for any distribution according to Chebyshev's theorem. For comparison, Figure~\ref{fig:case1} illustrates the original Bland-Altman bands and highlights that the decision cannot be solely based on the majority of points falling within the bands, as this is always the case. Additionally, these bands cannot provide information about any regression slope, as they are always horizontal. The correct approach is to use the hyperbolic-shaped 95\% confidence band, as shown in the precision tests at Figures~\ref{fig:case1},~\ref{fig:case2a},~\ref{fig:case2b},~\ref{fig:case2c},~and~\ref{fig:case3}, which allows for the assessment of the existence of a slope-zero line, considering the precision between the two measurement methods. These bands can be inclined depending on the precision relationship between the methods, thus leading to the rejection of the null hypothesis of precision equivalence.}

When one pursues the comparison of two techniques, non-rejection of the null hypothesis is not enough and the acceptance of equivalence (the acceptance of the null hypothesis) is conceptually necessary. Power computation obtained from a sample \textit{a posteriori} is meaningless~\cite{Gerard1998}, therefore planning of sample size along with study design \textit{a priori} is crucial to preserve statistical power. Budd et al.~\cite{Budd2013} propose at least 100 observations to claim consistency of a candidate measurement procedure applicable to different populations (item 6.3, page 12) down to 40 observations under more controlled laboratory conditions (item 7.2, page 15). However, this same source deals with more than a measure of each technique from the same patient with average or median, from which we disagree: it affects the computation of $\lambda$, wastes information, and, consequently, brings an ethical problem when invasive techniques are under assessment. Linnet also approached this issue, stating that sample sizes between 40 and 100 usually are to be reconsidered~\cite{Linnet1999}; the ideal number depends on the quotient between maximum and minimum measurements, proposing numbers from small sample sizes up to numbers in the order of 500 pairs of measurements (with mention to extreme numbers of thousands). Perhaps, some classic Bland and Altman examples applied here and many other published studies may be below the limit and allow only the rejection/non-rejection of null hypotheses without enough power to define true equivalence along the three statistical steps presented here.

In conclusion, it is possible to test whether two techniques may have full equivalence, preserving graphical communication according to Bland and Altman's principles, but adding robust and suitable inferential statistics. This approach decomposes the equivalence in accuracy, precision, and agreement for measurement techniques in such a way that, when full equivalence does not verify, this decomposition may help the location of the source of the problem in order to fix a new technique. Applications of the selected statistical methods using R provide automatized and standardized outputs of an otherwise complex calculation for better communication among researchers.

%% file: main.bbl
\begin{thebibliography}{48}
\providecommand{\natexlab}[1]{#1}
\providecommand{\url}[1]{\texttt{#1}}
\expandafter\ifx\csname urlstyle\endcsname\relax
  \providecommand{\doi}[1]{doi: #1}\else
  \providecommand{\doi}{doi: \begingroup \urlstyle{rm}\Url}\fi

\bibitem[Bland and Altman(1986)]{Bland1986}
J~Martin Bland and Douglas~G Altman.
\newblock {Statistical methods for assessing agreement between two methods of
  clinical measurement}.
\newblock \emph{The Lancet}, 327\penalty0 (8476):\penalty0 307--310, 6 1986.
\newblock ISSN 01406736.
\newblock \doi{10.1016/S0140-6736(86)90837-8}.

\bibitem[Pesola et~al.(2010)Pesola, Pesola, O'Donnell, Pesola, Chinchilli,
  Magari, and Saari]{Pesola2010}
Gene~R. Pesola, Gene~R. Pesola, Pamela O'Donnell, Helen~R. Pesola, Vernon~M.
  Chinchilli, Robert~T. Magari, and Arthur~F. Saari.
\newblock {Comparison of the ATS versus EU mini wright peak flow meter in
  normal volunteers}.
\newblock \emph{Journal of Asthma}, 47\penalty0 (10), 2010.
\newblock ISSN 02770903.
\newblock \doi{10.3109/02770903.2010.514639}.

\bibitem[Misyura et~al.(2018)Misyura, Sukhai, Kulasignam, Zhang, Kamel-Reid,
  and Stockley]{Misyura2018}
Maksym Misyura, Mahadeo~A. Sukhai, Vathany Kulasignam, Tong Zhang, Suzanne
  Kamel-Reid, and Tracy~L. Stockley.
\newblock {Improving validation methods for molecular diagnostics: Application
  of Bland-Altman, Deming and simple linear regression analyses in assay
  comparison and evaluation for next-generation sequencing}.
\newblock \emph{Journal of Clinical Pathology}, 71\penalty0 (2), 2018.
\newblock ISSN 14724146.
\newblock \doi{10.1136/jclinpath-2017-204520}.

\bibitem[Atkinson and Nevill(1998)]{Atkinson1998}
Greg Atkinson and Alan~M Nevill.
\newblock {Statistical methods for assessing measurement error (reliability) in
  variables relevant to sports medicine}.
\newblock \emph{Sports Medicine}, 26\penalty0 (4), 1998.
\newblock ISSN 01121642.
\newblock \doi{10.2165/00007256-199826040-00002}.

\bibitem[Shimada et~al.(2015)Shimada, Kario, Kushiro, Teramukai, Ishikawa,
  Kobayashi, and Saito]{Shimada2015}
Kazuyuki Shimada, Kazuomi Kario, Toshio Kushiro, Satoshi Teramukai, Yusuke
  Ishikawa, Fumiaki Kobayashi, and Ikuo Saito.
\newblock {Differences between clinic blood pressure and morning home blood
  pressure, as shown by Bland-Altman plots, in a large observational study
  (HONEST study)}.
\newblock \emph{Hypertension Research}, 38\penalty0 (12), 2015.
\newblock ISSN 13484214.
\newblock \doi{10.1038/hr.2015.88}.

\bibitem[Lo et~al.(2017)Lo, Zhao, Chen, Lei, Huang, and Tong]{Lo2017}
Wai Leung~Ambrose Lo, Jiang~Li Zhao, Ling Chen, Di~Lei, Dong~Feng Huang, and
  Kin~Fai Tong.
\newblock {Between-days intra-rater reliability with a hand held myotonometer
  to quantify muscle tone in the acute stroke population}.
\newblock \emph{Scientific Reports}, 7\penalty0 (1), 2017.
\newblock ISSN 20452322.
\newblock \doi{10.1038/s41598-017-14107-3}.

\bibitem[Aasvee et~al.(2015)Aasvee, Rasmussen, Kelly, Kurvinen, Giacchi, and
  Ahluwalia]{Aasvee2015}
Katrin Aasvee, Mette Rasmussen, Colette Kelly, Elvira Kurvinen,
  Mariano~Vincenzo Giacchi, and Namanjeet Ahluwalia.
\newblock {Validity of self-reported height and weight for estimating
  prevalence of overweight among Estonian adolescents: The Health Behaviour in
  School-aged Children study}.
\newblock \emph{BMC Research Notes}, 8\penalty0 (1), 2015.
\newblock ISSN 17560500.
\newblock \doi{10.1186/s13104-015-1587-9}.

\bibitem[Jones et~al.(2011)Jones, Dobson, and O'brian]{Jones2011}
M.~Jones, A.~Dobson, and S.~O'brian.
\newblock {A graphical method for assessing agreement with the mean between
  multiple observers using continuous measures}.
\newblock \emph{International Journal of Epidemiology}, 40\penalty0 (5), 2011.
\newblock ISSN 03005771.
\newblock \doi{10.1093/ije/dyr109}.

\bibitem[Taff{\'{e}} et~al.(2020)Taff{\'{e}}, Halfon, and Halfon]{Taffe2020}
Patrick Taff{\'{e}}, Patricia Halfon, and Matthieu Halfon.
\newblock {A new statistical methodology overcame the defects of the
  Bland–Altman method}.
\newblock \emph{Journal of Clinical Epidemiology}, 124, 2020.
\newblock ISSN 18785921.
\newblock \doi{10.1016/j.jclinepi.2020.03.018}.

\bibitem[Parker et~al.(2020)Parker, Scott, In{\'{a}}cio, and
  Stevens]{Parker2020}
Richard~A. Parker, Charles Scott, Vanda In{\'{a}}cio, and Nathaniel~T. Stevens.
\newblock {Using multiple agreement methods for continuous repeated measures
  data: A tutorial for practitioners}.
\newblock \emph{BMC Medical Research Methodology}, 20\penalty0 (1), 2020.
\newblock ISSN 14712288.
\newblock \doi{10.1186/s12874-020-01022-x}.

\bibitem[Creasy(1956)]{Creasy1956}
Monica~A. Creasy.
\newblock {Confidence Limits for the Gradient in the Linear Functional
  Relationship}.
\newblock \emph{Journal of the Royal Statistical Society: Series B
  (Methodological)}, 18\penalty0 (1), 1956.
\newblock ISSN 0035-9246.
\newblock \doi{10.1111/j.2517-6161.1956.tb00211.x}.

\bibitem[Zou(2013)]{Zou2013}
G.~Y. Zou.
\newblock {Confidence interval estimation for the Bland-Altman limits of
  agreement with multiple observations per individual}.
\newblock \emph{Statistical Methods in Medical Research}, 22\penalty0 (6),
  2013.
\newblock ISSN 09622802.
\newblock \doi{10.1177/0962280211402548}.

\bibitem[Carkeet(2015)]{Carkeet2015}
Andrew Carkeet.
\newblock {Exact parametric confidence intervals for bland-altman limits of
  agreement}.
\newblock \emph{Optometry and Vision Science}, 92\penalty0 (3), 2015.
\newblock ISSN 15389235.
\newblock \doi{10.1097/OPX.0000000000000513}.

\bibitem[Taff{\'{e}}(2020)]{Taffe2020b}
Patrick Taff{\'{e}}.
\newblock {Assessing bias, precision, and agreement in method comparison
  studies}.
\newblock \emph{Statistical Methods in Medical Research}, 29\penalty0 (3),
  2020.
\newblock ISSN 14770334.
\newblock \doi{10.1177/0962280219844535}.

\bibitem[Christensen et~al.(2020)Christensen, Borgbjerg, B{\o}rty, and
  B{\o}gsted]{Christensen2020}
Heidi~S. Christensen, Jens Borgbjerg, Lars B{\o}rty, and Martin B{\o}gsted.
\newblock {On Jones et al.’s method for extending Bland-Altman plots to
  limits of agreement with the mean for multiple observers}.
\newblock \emph{BMC Medical Research Methodology}, 20\penalty0 (1), 2020.
\newblock ISSN 14712288.
\newblock \doi{10.1186/s12874-020-01182-w}.

\bibitem[Watson and Petrie(2010)]{Watson2010}
P.~F. Watson and A.~Petrie.
\newblock {Method agreement analysis: A review of correct methodology}, 2010.
\newblock ISSN 0093691X.

\bibitem[Giavarina(2015)]{Giavarina2015}
Davide Giavarina.
\newblock {Understanding Bland Altman analysis}.
\newblock \emph{Biochemia Medica}, 25\penalty0 (2), 2015.
\newblock ISSN 13300962.
\newblock \doi{10.11613/BM.2015.015}.

\bibitem[Frost()]{Frost}
Jim Frost.
\newblock {Chebyshev’s Theorem in Statistics}.
\newblock URL
  \url{https://statisticsbyjim.com/basics/chebyshevs-theorem-in-statistics}.

\bibitem[Savage(1961)]{Savage1961}
Richard~I. Savage.
\newblock {Probability inequalities of the Tchebycheff type}.
\newblock \emph{Journal of Research of the National Bureau of Standards Section
  B Mathematics and Mathematical Physics}, 65B\penalty0 (3), 1961.
\newblock ISSN 0022-4340.
\newblock \doi{10.6028/jres.065b.020}.

\bibitem[Silveira and Siqueira(2021)]{eirasBA}
P~S~P Silveira and J~O Siqueira.
\newblock {R package: eirasBA}, 2021.
\newblock URL
  \url{https://dataverse.harvard.edu/dataset.xhtml?persistentId=doi:10.7910/DVN/AGJPZH}.

\bibitem[Isaac(1970)]{Isaac1970}
Paul~D. Isaac.
\newblock {Linear regression, structural relations, and measurement error}.
\newblock \emph{Psychological Bulletin}, 74\penalty0 (3), 1970.
\newblock ISSN 00332909.
\newblock \doi{10.1037/h0029777}.

\bibitem[Thoresen and Laake(2007)]{Thoresen2007}
Magne Thoresen and Petter Laake.
\newblock {On the simple linear regression model with correlated measurement
  errors}.
\newblock \emph{Journal of Statistical Planning and Inference}, 137\penalty0
  (1), 2007.
\newblock ISSN 03783758.
\newblock \doi{10.1016/j.jspi.2005.09.001}.

\bibitem[Hedberg and Ayers(2015)]{Hedberg2015}
E.~C. Hedberg and Stephanie Ayers.
\newblock {The power of a paired t-test with a covariate}.
\newblock \emph{Social Science Research}, 50, 2015.
\newblock ISSN 0049089X.
\newblock \doi{10.1016/j.ssresearch.2014.12.004}.

\bibitem[Shukla(1973)]{Shukla1973}
G.~K. Shukla.
\newblock {Some Exact Tests of Hypotheses about Grubbs's Estimators}.
\newblock \emph{Biometrics}, 29\penalty0 (2):\penalty0 373, 6 1973.
\newblock ISSN 0006341X.
\newblock \doi{10.2307/2529399}.

\bibitem[Glaister(2001)]{Glaister2001}
P.~Glaister.
\newblock {85.13 Least squares revisited}.
\newblock \emph{The Mathematical Gazette}, 85\penalty0 (502), 2001.
\newblock ISSN 0025-5572.
\newblock \doi{10.2307/3620485}.

\bibitem[Oldham(1962)]{Oldham1962}
P.~D. Oldham.
\newblock {A note on the analysis of repeated measurements of the same
  subjects}.
\newblock \emph{Journal of Chronic Diseases}, 15\penalty0 (10), 1962.
\newblock ISSN 00219681.
\newblock \doi{10.1016/0021-9681(62)90116-9}.

\bibitem[Linnet(1998)]{Linnet1998}
Kristian Linnet.
\newblock {Performance of Deming regression analysis in case of misspecified
  analytical error ratio in method comparison studies}.
\newblock \emph{Clinical Chemistry}, 44\penalty0 (5), 1998.
\newblock ISSN 00099147.
\newblock \doi{10.1093/clinchem/44.5.1024}.

\bibitem[Kummell(1879)]{Kummell1879}
Chas.~H. Kummell.
\newblock {Reduction of Observation Equations Which Contain More Than One
  Observed Quantity}.
\newblock \emph{The Analyst}, 6\penalty0 (4), 1879.
\newblock ISSN 07417918.
\newblock \doi{10.2307/2635646}.

\bibitem[Albert(1991)]{Albert1992}
Adelin Albert.
\newblock \emph{{Statistical methods in laboratory medicine.}}, volume~11.
\newblock Butterworth-Heinemann Ltd., Oxford,, 1991.
\newblock \doi{10.1002/sim.4780111315}.

\bibitem[Efron(2007)]{Efron2007}
B.~Efron.
\newblock {Bootstrap Methods: Another Look at the Jackknife}.
\newblock \emph{The Annals of Statistics}, 7\penalty0 (1), 2007.
\newblock ISSN 0090-5364.
\newblock \doi{10.1214/aos/1176344552}.

\bibitem[Shoukri(2010)]{Shoukri2010}
Mohamed~M. Shoukri.
\newblock \emph{{Measures of Interobserver Agreement and Reliability}}.
\newblock 2010.
\newblock \doi{10.1201/b10433}.

\bibitem[{NCSS Statistical Software [book on the Internet]}()]{NCSS}
{NCSS Statistical Software [book on the Internet]}.
\newblock {Deming regression}.
\newblock URL
  \url{https://ncss-wpengine.netdna-ssl.com/wp-content/themes/ncss/pdf/Procedures/NCSS/Deming\_Regression.pdf}.

\bibitem[Antonakis et~al.(2010)Antonakis, Bendahan, Jacquart, and
  Lalive]{Antonakis2010}
John Antonakis, Samuel Bendahan, Philippe Jacquart, and Rafael Lalive.
\newblock {On making causal claims: A review and recommendations}, 2010.
\newblock ISSN 10489843.

\bibitem[McCartin(2003)]{McCartin2003}
Brian~J. McCartin.
\newblock {A geometric characterization of linear regression}.
\newblock \emph{Statistics}, 37\penalty0 (2), 2003.
\newblock ISSN 02331888.
\newblock \doi{10.1080/0223188031000112881}.

\bibitem[Roberts(2012)]{Roberts2012}
Steven Roberts.
\newblock {Statistical Thinking in Epidemiology. By Y.-K. Tu and M. Gilthorpe.
  Boca Raton, Florida: CRC Press. 2011. 231 pages. UK{\pounds}57.99 (hardback).
  ISBN 978-1-4200-9991-1.}
\newblock \emph{Australian {\&} New Zealand Journal of Statistics}, 54\penalty0
  (4), 2012.
\newblock ISSN 1369-1473.
\newblock \doi{10.1111/j.1467-842x.2012.00675.x}.

\bibitem[Anscombe(1973)]{Anscombe1973}
F.~J. Anscombe.
\newblock {Graphs in statistical analysis}.
\newblock \emph{American Statistician}, 27\penalty0 (1):\penalty0 17--21, 1973.
\newblock ISSN 15372731.
\newblock \doi{10.1080/00031305.1973.10478966}.

\bibitem[Bland and Altman(1999)]{Bland1999}
J~Martin Bland and Douglas~G Altman.
\newblock {Measuring agreement in method comparison studies}.
\newblock \emph{Statistical Methods in Medical Research}, 8\penalty0 (2), 1999.
\newblock ISSN 0962-2802.
\newblock \doi{10.1177/096228029900800204}.

\bibitem[Videira and Vieira(2011)]{Videira2011}
Rogerio~L.R. Videira and Joaquim~E. Vieira.
\newblock {What rules of thumb do clinicians use to decide whether to
  antagonize nondepolarizing neuromuscular blocking drugs?}
\newblock \emph{Anesthesia and Analgesia}, 113\penalty0 (5), 2011.
\newblock ISSN 00032999.
\newblock \doi{10.1213/ANE.0b013e31822c986e}.

\bibitem[Altman and Bland(1983)]{Altman1983}
D~G Altman and J~M Bland.
\newblock {JSTOR: Journal of the Royal Statistical Society. Series D (The
  Statistician), Vol. 32, No. 3 (Sep., 1983), pp. 307-317}.
\newblock \emph{The statistician}, 1983.

\bibitem[Altman and Bland(1986)]{Altman1986}
D.~G. Altman and J.~M. Bland.
\newblock {Comparison of methods of measuring blood pressure.}, 1986.
\newblock ISSN 14702738.

\bibitem[Bland and Altman(2003)]{Bland2003}
J~M Bland and D~G Altman.
\newblock {Applying the right statistics: Analyses of measurement studies},
  2003.
\newblock ISSN 09607692.

\bibitem[Datta(2017)]{blandr}
Deepankar Datta.
\newblock {blandr: a Bland-Altman Method Comparison package for R}, 2017.
\newblock URL \url{https://github.com/deepankardatta/blandr}.

\bibitem[Peng et~al.(2022)Peng, Taff{\'{e}}, and Williamson]{MethodCompare}
Mingkai Peng, Patrick Taff{\'{e}}, and Tyler Williamson.
\newblock {MethodCompare: Bias and Precision Plots}, 2022.
\newblock URL \url{https://CRAN.R-project.org/package=MethodCompare}.

\bibitem[Carstensen et~al.(2020)Carstensen, Gurrin, Ekstr{\o}m, and
  Figurski]{MethComp}
Bendix Carstensen, Lyle Gurrin, Claus~Thorn Ekstr{\o}m, and Michal Figurski.
\newblock {MethComp: Analysis of Agreement in Method Comparison Studies}, 2020.
\newblock URL \url{https://CRAN.R-project.org/package=MethComp}.

\bibitem[Potapov et~al.(2023)Potapov, Model, Schuetzenmeister, Manuilova,
  Dufey, and Raymaekers]{mcr}
Sergej Potapov, Fabian Model, Andre Schuetzenmeister, Ekaterina Manuilova,
  Florian Dufey, and Jakob Raymaekers.
\newblock {mcr: Method Comparison Regression}, 2023.
\newblock URL \url{https://CRAN.R-project.org/package=mcr}.

\bibitem[Gerard et~al.(1998)Gerard, Smith, and Weerakkody]{Gerard1998}
Patrick~D. Gerard, David~R. Smith, and Govinda Weerakkody.
\newblock {Limits of Retrospective Power Analysis}.
\newblock \emph{The Journal of Wildlife Management}, 62\penalty0 (2), 1998.
\newblock ISSN 0022541X.
\newblock \doi{10.2307/3802357}.

\bibitem[Budd et~al.(2013)Budd, Durham, Gwise, Iriarte, Kallner, Linnet,
  .Magari, and Vaks]{Budd2013}
Jeffrey~R. Budd, A.P. Durham, T.E. Gwise, B.~Iriarte, A.~Kallner, K.~Linnet,
  R~.Magari, and J.E. Vaks, editors.
\newblock \emph{{EP09A3-Measurement Procedure Comparison and Bias Estimation
  Using Patient Samples: approved guideline}}, volume~11.
\newblock The Clinical and Laboratory Standards Institute, Pittsburgh, PA, US,
  3 edition, 2013.
\newblock ISBN 9781562388881.

\bibitem[Linnet(1999)]{Linnet1999}
Kristian Linnet.
\newblock {Necessary sample size for method comparison studies based on
  regression analysis}.
\newblock \emph{Clinical Chemistry}, 45\penalty0 (6 I), 1999.
\newblock ISSN 00099147.
\newblock \doi{10.1093/clinchem/45.6.882}.

\end{thebibliography}
